\documentclass[aps,prl,reprint,superscriptaddress,preprintnumbers,amsmath,nofootinbib]{revtex4-2}
\usepackage{graphicx}% Include figure files
\usepackage{dcolumn}% Align table columns on decimal point
\usepackage{bm}% bold math
\usepackage{braket}
\usepackage{epstopdf}
\usepackage{hyperref}% add hypertext capabilities
\usepackage{color}
\usepackage{cancel}
\usepackage{amsfonts}
\begin{document}
\preprint{RIKEN-iTHEMS-Report-25, FQSP-25-2, YITP-25-109}

%Title of paper
\title{Decoding Two-Particle States in QCD with Spatial Wavefunctions}
%=======================================================================================

\author{Yan Lyu}
\email{yan.lyu@riken.jp}
\affiliation{RIKEN Center for Interdisciplinary Theoretical and
Mathematical Sciences (iTHEMS), RIKEN, Wako 351-0198, Japan}

\author{Sinya Aoki}
\email{saoki@yukawa.kyoto-u.ac.jp}
\affiliation{Fundamental Quantum Science Program (FQSP), TRIP Headquarters, RIKEN, Wako 351-0198, Japan}
\affiliation{Center for Gravitational Physics and Quantum Information, Yukawa Institute for Theoretical Physics, Kyoto University, Kyoto 606-8502, Japan}

\author{Takumi Doi}
\email{doi@ribf.riken.jp}
\affiliation{RIKEN Center for Interdisciplinary Theoretical and
Mathematical Sciences (iTHEMS), RIKEN, Wako 351-0198, Japan}

\author{Tetsuo Hatsuda}
\email{thatsuda@riken.jp}
\affiliation{RIKEN Center for Interdisciplinary Theoretical and
Mathematical Sciences (iTHEMS), RIKEN, Wako 351-0198, Japan}
\affiliation{Kavli Institute for the Physics and Mathematics of the Universe (Kavli IPMU), WPI, The University of Tokyo, Kashiwa, Chiba 277-8568, Japan}

\author{Kotaro Murakami}
\email{kotaro.murakami@yukawa.kyoto-u.ac.jp}
\affiliation{Department of Physics, Institute of Science Tokyo, 2-12-1 Ookayama, Megro, Tokyo 152-8551, Japan}
\affiliation{RIKEN Center for Interdisciplinary Theoretical and
Mathematical Sciences (iTHEMS), RIKEN, Wako 351-0198, Japan}

\author{Takuya Sugiura}
\email{sugiura@rcnp.osaka-u.ac.jp}
\affiliation{Faculty of Data Science, Rissho University, Kumagaya 360-0194, Japan}
\affiliation{RIKEN Center for Interdisciplinary Theoretical and
Mathematical Sciences (iTHEMS), RIKEN, Wako 351-0198, Japan}

\date{\today}
%=======================================================================================
\begin{abstract}
A systematic framework for constructing optimized interpolating operators strongly coupled to QCD two-particle states is developed,
which is achieved by incorporating inter-hadron spatial wavefunctions.
To efficiently implement these operators in lattice QCD,
a novel quark smearing technique utilizing noise vectors is proposed.
Applied to the $\Omega_{ccc}\Omega_{ccc}$ system, 
these optimized operators prove superior to combinations of limited plane-wave operators,
enabling the resolution of distinct eigenstates separated by only $\sim 5$ MeV near the threshold $2m_{\Omega_{ccc}} \simeq 9700$ MeV.
This exceptional resolving power opens new possibilities for studies of a wide range of hadronic systems in QCD.
\end{abstract}
%=======================================================================================

%\keywords{}

%\maketitle must follow title, authors, abstract, \pacs, and \keywords
\maketitle

%=================================================

{\it Introduction.$-$} Determining the eigenstates of strongly interacting systems remains a formidable challenge across many areas of physics, from condensed matter to particle physics. In particular, in Quantum Chromodynamics (QCD), the fundamental theory of the strong interaction, color confinement prevents the colored quarks and gluons that appear in the Lagrangian from existing as isolated states. Instead, they self-organize into color-neutral 
hadrons (mesons and baryons) as well as their bound and scattering states.
 Moreover, in the course of forming such bound and scattering states, nontrivial spatial structures can emerge, giving rise to familiar nuclear-physics phenomena such as the quadrupole-deformed deuteron and the diverse shapes of heavy nuclei~\cite{Meng2016}. Understanding these spatial structures reflects the hadron–hadron interaction, including its range and partial-wave content, and provides essential information not only on the energy spectrum of the eigenstates but also on quantum transitions between them.

\begin{figure}[htbp]
  \centering
  \includegraphics[width=8.0cm]{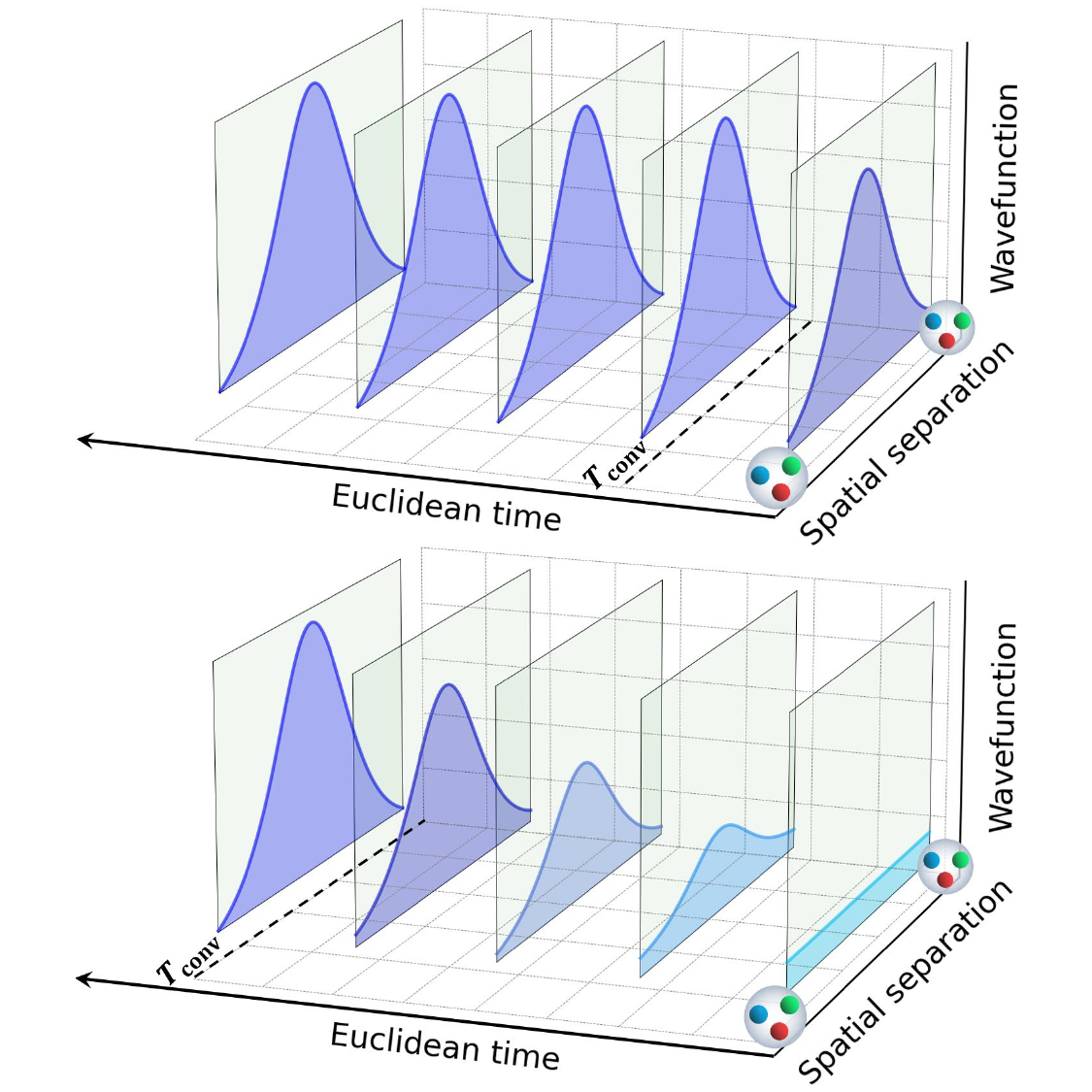}
  \caption{Sketch of Euclidean-time evolution of two-hadron wavefunctions.
  A realistic initial wavefunction (top) converges to the asymptotic form quickly, while a flat one (bottom) requires a longer time $T_{\mathrm{conv}}$ for convergence.
  The convergence time slice is indicated by dashed lines.
  } \label{Fig_evolution}
\end{figure}

Direct access to such spatial information from first-principles QCD calculations requires the isolation of targeted states, 
a task that remains challenging even for simple two-particle systems.
Despite successes in the single-hadron sector~\cite{FLAG2024}, the lattice formulation faces a greater yet essential challenge in isolating two-hadron eigenstates for determining their spectra and matrix elements~\cite{Drischler2019}. 
The difficulty arises because the energy gaps between adjacent two-hadron states are kinematically determined, scaling as $\Delta E^{\text{el}} \sim \dfrac{4\pi^2}{m_{\rm H} L^2}$ and therefore decrease with lattice size $L$ and hadron mass $m_{\rm H}$. 
This contrasts with single-hadron states, whose gaps are set by intrinsic scales insensitive to $L$, such as the pion mass.
The resulting dense spectrum of two-hadron states necessitates highly optimized operators that efficiently couple to and rapidly isolate the targeted states, thereby mitigating the exponential degradation of the signal at large Euclidean times~\cite{Parisi:1983ae, Lepage1989}.
As sketched in Fig.~\ref{Fig_evolution}, an initial state with realistic spatial structure
converges more rapidly under Euclidean-time evolution $e^{-\hat Ht}|\psi\rangle$.
Previous efforts in this direction include extended two-hadron operators~\cite{Kurth:2013tua,Berkowitz:2015eaa, Wu:2021xvz, Amarasinghe:2021lqa} and 
plane-wave based variational studies\cite{Luscher:1990ck, Blossier:2009kd,Francis2019,Amarasinghe:2021lqa,BaSc:2025yhy}.

%==========================
%==========================

In this Letter, we present a systematic framework for constructing optimized two-hadron operators by incorporating realistic spatial wavefunctions. 
We introduce a novel quark-smearing technique that enables the efficient implementation of such operators. 
In a test case, the optimized operators exhibit substantially improved overlap with target states compared to those built from a limited set of plane-wave operators, yielding characteristic spatial structures for both bound and scattering states and resolving nearby energy levels separated by as little as $\sim5$ MeV.
This unprecedented resolving power for QCD eigenstates opens new opportunities for studying the structure and dynamics of multi-hadron systems.
The full technical details and extended discussion are presented in the companion Regular Article~\cite{Lyu:2025ncq}.

%=================================================
{\it Optimized two-hadron operators.$-$}
As a representative example, let us consider a system of two identical baryons in the CM frame within a finite box $V$ and the lattice sites $\Lambda$.
The four-point baryon correlation function $\mathcal{F}$ for this system can be generally 
written as, 
\begin{align}\label{Eq-R}
    &\mathcal{F}(\vec r_\text{snk}, t;\vec r_\text{src}, t_0) \nonumber \\ 
   &=\frac{1}{V^2}\sum_{\vec{x},\vec{y}\in \Lambda}\langle   B(\vec x+\vec r_\text{snk},t)  B(\vec x,t)  
  \bar{B}(\vec y+\vec r_\text{src},t_0) \bar{B}(\vec y,t_0)\rangle \nonumber \\
  &=\sum^M_{n=0} \psi_n(\vec r_\text{snk}) \psi^*_n(\vec r_\text{src}) e^{-E_n(t-t_0)} + O(e^{- E^*(t-t_0)}). 
\end{align}
Here, $B(\vec x, t)$ denotes a local interpolating operator for a baryon of mass $m_B$. The 
product of two baryon operators at the source and  sink  is assumed to be projected onto a specific spin and isospin channel.
Summations over the spatial indices $\vec{x}$ (sink) and $\vec{y}$ (source) are carried out over all lattice sites $\Lambda$. 
The second line of Eq.~(\ref{Eq-R}) is obtained by inserting complete physical states with baryon number 2.
In particular,   any possible bound state(s) as well as the elastic scattering states of two baryons below the inelastic threshold $E^*$ are labeled by discrete energies $E_n (< E^*)$ with  $n=0, 1, \cdots, M$.
These states are characterized by the  equal-time Nambu-Bethe-Salpeter (NBS) amplitude,
\begin{align}
    \psi_n(\vec r)=\frac{1}{V}\sum_{\vec x\in\Lambda} \langle 0| B(\vec x+\vec r,0) B(\vec x,0)|E_n\rangle ,\label{Eq-NBS}    
\end{align}
which represents the probability amplitude of locating one baryon at $\vec{x}$ and another at $\vec{x} + \vec{r}$ from the state $|E_n\rangle$.  It serves as a generalized form of the relative stationary wavefunction for composite particles in quantum field theory~\cite{Ishii2007}.
The contribution from elastic or bound states becomes dominant over inelastic states if the time separation between the sink and source is sufficiently large, $t - t_0 \gg {(\Delta E^*)}^{-1}$, where $\Delta E^* = E^* - 2m_B$ represents the energy gap to the inelastic channel.

The states $|E_n \rangle $ are orthonormal with each other since they are the exact eigenstates of the QCD Hamiltonian, while $\psi_n(\vec r)$ are not necessary so, since they contain only the partial information of $|E_n \rangle $.
Let us define the dual basis $\Psi_n(\vec r) $ with the norm kernel ${\cal K}$ as,
\begin{align}
\Psi_n(\vec r)  = \sum_{n'=0}^M \mathcal\psi_{n'}(\vec r){K}^{-1}_{n'n},\label{Eq-APsi} \ \ 
\mathcal{K}_{n'n}& = \langle \psi_{n'}| \psi_{n} \rangle.
\end{align}
The dual basis satisfies $\langle \Psi_n| \psi_{n'} \rangle = \frac{1}{V}\sum\limits_{\vec r\in\Lambda} \Psi^*_n(\vec r) \psi_{n'}(\vec r) = \delta_{nn'}$, which allow us to define a set of highly optimized two-baryon operators $\{O_n\}$ for specific states as,
\begin{align}\label{Eq-O}
    O_n(t) = \frac{1}{V^2}\sum_{\vec x,\vec r\in\Lambda}  B(\vec x+\vec r,t) B(\vec x,t) \Psi^*_n(\vec r).
\end{align}
Applying such operators at the source and sink is equivalent to projecting the correlation function in Eq.~(\ref{Eq-R}) onto specific states, yielding:
\begin{align}
    R_n(\vec r_\text{snk},t)  &=\frac{1}{V}\sum_{\vec r_\text{src}\in\Lambda}\mathcal{F}(\vec r_\text{snk}, t;\vec r_\text{src}, 0) 
    \Psi_n(\vec r_\text{src}) / e^{-2m_Bt}
    \nonumber \\
    & = \psi_n(\vec r_\text{snk}) e^{-\Delta E_nt}, \label{Eq-Rr} \\
    R_n(t)  &=\frac{1}{V}\sum_{\vec r_\text{snk}\in\Lambda}\Psi^*_n(\vec r_\text{snk}) R_n(\vec r_\text{snk},t)\nonumber \\
    & = e^{-\Delta E_n t},  \label{Eq-Rt}
\end{align}
where $\Delta E_n = E_n-2m_B$ and the contribution from inelastic states is neglected by assuming $t \gg (\Delta E^*)^{-1} $.

Although it is generally difficult to obtain $\psi_{n\le M}(\vec r)$, 
we can extract their approximate form in the following procedure as proposed in our previous studies~\cite{Iritani2019Jhep, Lyu2022}:
We first calculate the leading-order local potential using the time-dependent HAL QCD method\cite{Ishii2007,Ishii2012}, 
which enables the extraction of a potential that effectively describes scattering properties from a lattice correlation function (not necessarily dominated by a single state) as follows.
\begin{align}
    & V^{(0)}(\vec r) = \frac{1}{R^{(0)}(\vec r, t)}\left(\frac{1}{4m_B}\frac{\partial^2}{\partial t^2} - \frac{\partial}{\partial t} +\frac{\nabla^2}{m_B}\right) R^{(0)}(\vec r, t), \label{Eq-V} 
\end{align}
where $R^{(0)}(\vec r, t)$ is the correlation function obtained from 
some given (unoptimized) source operators (such as the wall source operator, where the quark field is projected to zero momentum).
We then solve the eigenvalue problem  for the Sch\"odinger equation defined on a finite box with this potential as, 
\begin{align}
    & \left[-\frac{\nabla^2}{m_B} +V^{(0)}(\vec r)\right]\psi^{(0)}_n (\vec r) = \frac{k^2_n}{m_B}\psi^{(0)}_n (\vec r),\label{Eq-FV}
\end{align}
where $\frac{k^2_n}{m_B}$ and $\psi^{(0)}_n$ are the eigenvalue and eigenfunction, respectively, and the corresponding energy is given by $\varepsilon_n=2\left(\sqrt{m_B^2+k^2_n}-m_B\right)$.
To improve the accuracy on $\psi_{n}(\vec r)$, we adopt the following iterative procedure.
Starting with the initial wavefunctions ${\psi^{(0)}_n(\vec r)}$, we compute the corresponding dual functions ${\Psi^{(0)}_n(\vec r)}$ using Eq.~(\ref{Eq-APsi}). These are then inserted into Eq.~(\ref{Eq-Rr}) to obtain the set ${R^{(1)}_n(\vec r, t)}$, along with the corresponding potentials $V_n^{(1)}(\vec r)$. 
These potentials are subsequently used in the Schr\"odinger equation to calculate the updated wavefunctions ${\psi^{(1)}_n(\vec r)}$.
The entire process can be repeated iteratively until convergence is achieved—signaled by the stabilization of the profiles ${R_n(\vec r, t)}$ with respect to time $t$, or equivalently, by the stability of the effective energies, $\Delta E^{\text{eff}}_n(t)=\ln\left[\frac{R_n(t)}{R_n(t+1)}\right]$, as defined from Eq.~\eqref{Eq-Rt}.

%=================================================
%==========================
\begin{figure}[htbp]
  \centering
  \includegraphics[width=8cm]{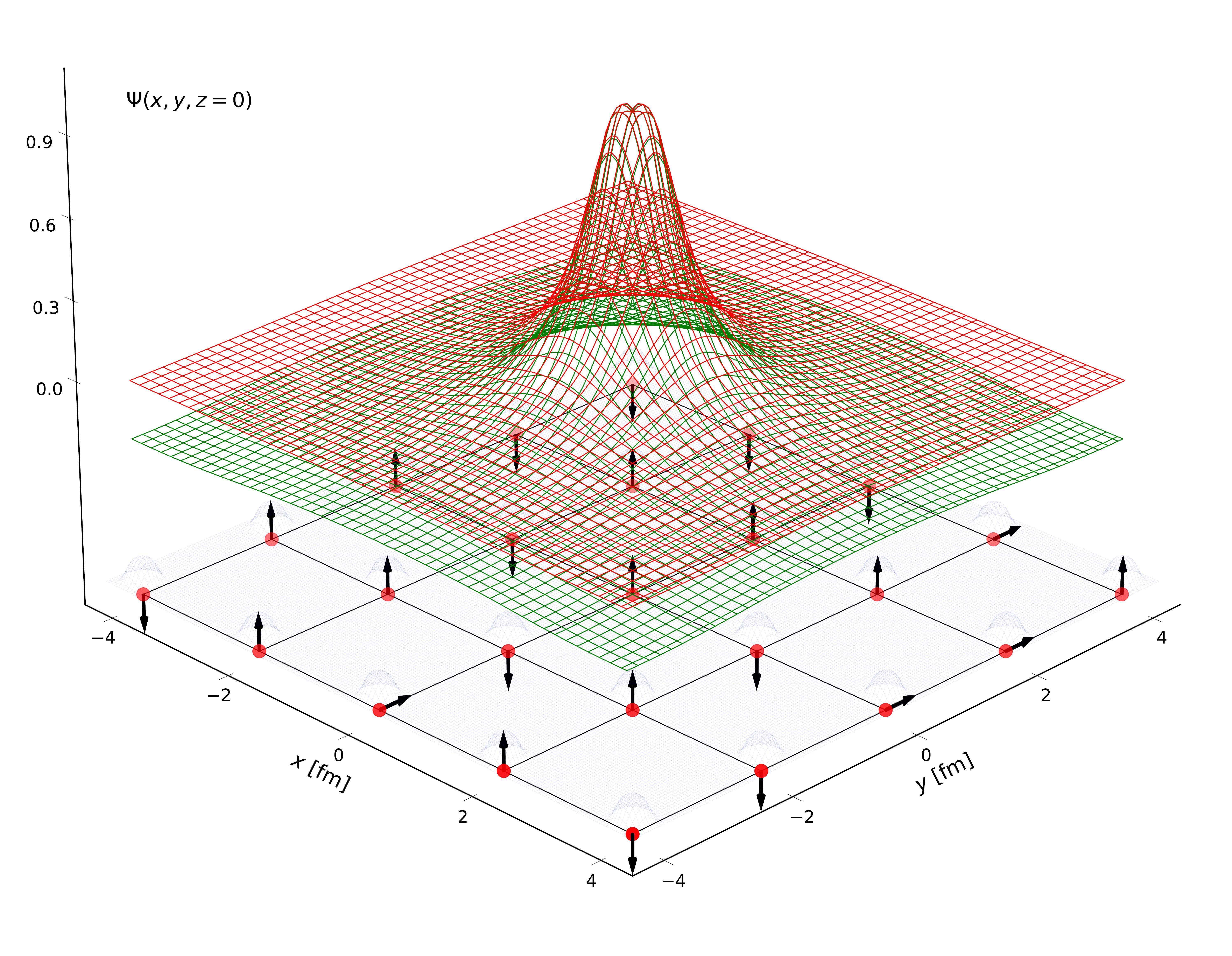}
  \caption{The novel quark smearing in Eq.~(\ref{Eq-Smear-2}). 
  Each of the multiple supports distributed on a sparse lattice grid (red points) contains a local smearing function $f(\vec r)$ (blue) fine-tuned for single hadron together with an independent $Z_3$ noise vector (arrows), and is weighted according to the cubic root of inter-hadron spatial wavefunctions $\Psi_0(\vec r)$ (red) and $\Psi_1(\vec r)$ (green), where $\Psi_{0,1}(\vec r)$ are normalized by $\Psi_{0,1}(\vec 0)$.
  } \label{Fig-smear}
\end{figure}
%==========================

{\it Novel quark smearing for source operators.$-$}
To implement the optimized operators Eq.~(\ref{Eq-O}) into the correlation functions in Eq.~(\ref{Eq-Rt}), it is necessary to compute the quark propagator $\langle  q(\vec x, t)\bar{ q}(\vec y,0)\rangle$ for all $\vec x,\vec y\in\Lambda$ (so called ``all-to-all'' propagators), such that the spatial wavefunction $\Psi_n(\vec r)$ can be faithfully incorporated between two local baryon operators at both the source and sink.
In lattice QCD, the quark propagator is typically computed using a smeared quark field $q_f$ at the source, $\langle q(\vec x, t)\bar{q}_f(\vec y, 0)\rangle=\sum\limits_{\vec r\in\Lambda}\langle q(\vec x, t)\bar{ q}(\vec r,0)\rangle f(\vec r-\vec y)$ for all $\vec x\in\Lambda$ and fixed $\vec y$, where $f(\vec r)$ is a (local) smearing function (such as the delta function, the exponential/Gaussian form,  etc.).
To avoid the prohibitive computational cost of computing the quark propagator for all spatial points $(\vec y)$ at the source, one can instead design appropriate source smearing functions that embed the wavefunction $\Psi_n(\vec r)$ directly.

In order to do so, we introduce the following two smearing functions,
\begin{align}
   & G(\vec r) = f(\vec r), \label{Eq-Smear-1}\\
   & F_n(\vec r) = \frac{1}{V^{1/3}_\text{sub}}\sum_{\vec r_0\in\Lambda_\text{sub}} Z_3(\vec r_0) \Psi^{1/3}_n(\vec r_0) f(\vec r - \vec r_0)\label{Eq-Smear-2},
\end{align}
where the local smearing function $f(\vec r)$ is fine-tuned to enhance coupling to the ground state of single baryon.
This construction gives $G(\vec r)$ a single support at the origin, while $F_n(\vec r)$ has multiple supports distributed on a sub lattice $\Lambda_\text{sub}$ with volume $V_\text{sub}$, each weighted according to the wavefunction factor $\Psi^{1/3}_n(\vec r_0)$ and rotated by an independent noise vector $Z_3\in \{1, e^{i2/3\pi}, e^{i4/3\pi}\}$ (see Fig.~\ref{Fig-smear}).
The $Z_3$ noise ensures that when combining three quark operators smeared with $F_n(\vec r)$, only terms from identical supports survive while cross terms between different supports vanish statistically under the $Z_3$ noise average. 
Thus, the optimized source operator $\bar{O}_n$ is achieved by combing three quark operators smeared with $G(\vec r)$ and another three smeared with $F_n(\vec r)$, 
\begin{align}
    (\bar{q}_G)^3(\bar{q}_{F_n})^3 = &\frac{1}{V_\text{sub}}\bar{B}(\vec 0)\sum_{\vec r_0\in\Lambda_\text{sub}}\Psi_n(\vec r_0) \bar{B}(\vec r_0)\nonumber\\
    \simeq& \frac{1}{V}\bar{B}(\vec 0)\sum_{\vec r_0\in\Lambda}\Psi_n(\vec r_0) \bar{B}(\vec r_0)~\label{Eq-sp-src},
\end{align}
where the smeared quark operators are defined as $\bar{q}_G=\sum\limits_{\vec r\in\Lambda} \bar{q}(\vec r) G(\vec r)$, similar for $\bar{q}_{F_n}$, and are assumed to be suitably combined in color and spinor spaces to form two baryon operators. 
The sub lattice $\Lambda_\text{sub}$ is defined through sparsening the original lattice grid $\Lambda$ (of size $L$) with an interval $l$,
\begin{align}
    \Lambda_\text{sub}=\{l\vec r~|~\vec r\in \mathbb{Z}^3, 0\leq r_{x,y,z}<L/l \}.
\end{align}
The difference between summing over $\Lambda_\text{sub}$ and $\Lambda$ in Eq.~(\ref{Eq-sp-src}) arises from high-momentum contamination (e.g., $p=2\pi/l$) and is suppressed for $t\gg\frac{m_Bl^2}{4\pi}$, as discussed in Refs.~\cite{Li:2020hbj,Detmold:2019fbk}.
In practice, $\Lambda_\text{sub}$ should be fine enough
for a good approximation while simultaneously sparse enough to reduce statistical fluctuations of $Z_3$ noises. The reference coordinate $\vec 0$ in Eq.~(\ref{Eq-sp-src}) should be randomly shifted for each configuration to ensure zero total momentum at the source.

%=================================================
%==========================
\begin{table}[htbp]
\caption{Time evolution of the spatial profile of correlation functions $R(\vec r, t)$, quantified by the residue factor
$\mathcal{L}[R(\vec r,t)]=\frac1V\sum\limits_{\vec r\in\Lambda}\left|\frac{R(\vec r, t)/R(\vec 0, t) }{R(\vec r, t_\text{f})/R(\vec 0, t_\text{f})} - 1\right|$ with $t_\text{f}/a=30$. }
\begin{tabular}{lccc}
  \hline\hline
    $t/a$ ~~~~~&$\mathcal{L}[R^{(0)}(\vec r, t)]$~~~~~&$\mathcal{L}[R^{(1)}_0(\vec r, t)]$~~~~~&$\mathcal{L}[R^{(1)}_1(\vec r, t)]$\\
  \hline
$20$  ~~~~~&$24.1(0.3)\%$~~~~~&$3.7(1.4)\%$~~~~~&$1.6(2.5)\%$ \\
$24$  ~~~~~&$12.5(0.2)\%$~~~~~&$1.4(1.1)\%$~~~~~&$0.8(1.0)\%$ \\
$28$  ~~~~~&$~3.6(0.1)\%$~~~~~&$0.5(0.5)\%$~~~~~&$0.4(0.4)\%$ \\
 \hline\hline
\end{tabular} \label{tab-t-dep}
\end{table}
%==========================                                                                                                            
{\it Application to $\Omega_{ccc}\Omega_{ccc}$.$-$}
For demonstration purposes, we consider the $\Omega_{ccc}\Omega_{ccc}$ system in $^1S_0$ channel for three reasons: (i) the heavy $\Omega_{ccc}$ results in dense spectra of two baryons  
on a large lattice box, necessitating  highly optimized operators for the state identification;
(ii) the previous lattice calculation~\cite{Lyu2021} has shown that this system supports a bound state, allowing us to test our method on both bound and scattering states; 
and (iii) the system exhibits smaller statistical fluctuations than lighter baryons, making optimized operator effects more evident.

The calculations are performed with HAL-conf-2023 lattice QCD gauge configurations, generated with ($2+1$) flavor quarks on a large box ($L=96, La\simeq8.1$ fm) at physical point ($m_\pi\simeq137$ MeV) by the HAL QCD Collaboration using the supercomputer Fugaku at RIKEN~\cite{Aoyama:2024cko}.
The relativistic heavy quark action~\cite{Aoki2003,Namekawa2017} is employed for charm quark ($m_{\Omega_{ccc}}\simeq4847$ MeV).
A local baryon operator $\Omega_{ccc}(x)=c^T(x)\mathcal{C}\gamma_k c(x) c(x)$ is adopted at the sink,
while a similar one with $c(x)$ being slightly smeared with an exponential function $f(\vec r)= e^{-B|\vec r|}$ with $B=0.475~a^{-1}$
fine-tuned for single $\Omega_{ccc}$ is used at the source with the Coulomb gauge fixing, to reduce statistic errors.

As suggested before, we first adopt the wall source to calculate $R^{(0)}(\vec r, t)$ to initiate the iterative process,
which of course receives contributions from multiple states and does not exhibit a stable profile against $t$,
as quantified by the residue factor
$\mathcal{L}[R^{(0)}(\vec r,t)]=\frac1V\sum\limits_{\vec r\in\Lambda}\left|\frac{R^{(0)}(\vec r, t)/R^{(0)}(\vec 0, t) }{R^{(0)}(\vec r, t_\text{f})/R^{(0)}(\vec 0, t_\text{f})} - 1\right|$ with $t_\text{f}/a=30$
shown in Table~\ref{tab-t-dep}.
On the other hand, the potential $V^{(0)}(\vec r, t)$ calculated via Eq.~(\ref{Eq-V}) is rather stable, thank to 
the fact that (elastic) state identification is not required in the time-dependent HAL QCD method~\cite{Ishii2012}.
Using the (orthogonal) wavefunctions $\{\psi^{(0)}_n(\vec r)\}$ of Eq.~(\ref{Eq-FV}) with the potential $V^{(0)}(\vec r)$ at $t/a=25$, the dual function $\{\Psi^{(0)}_n(\vec r)\}$ is constructed via Eq.~(\ref{Eq-APsi}). The first two dual functions $\Psi^{(0)}_{0,1}(\vec r)$ are shown in Fig.~\ref{Fig-smear}.

\begin{figure}[htbp]
  \centering
  \includegraphics[width=8.cm]{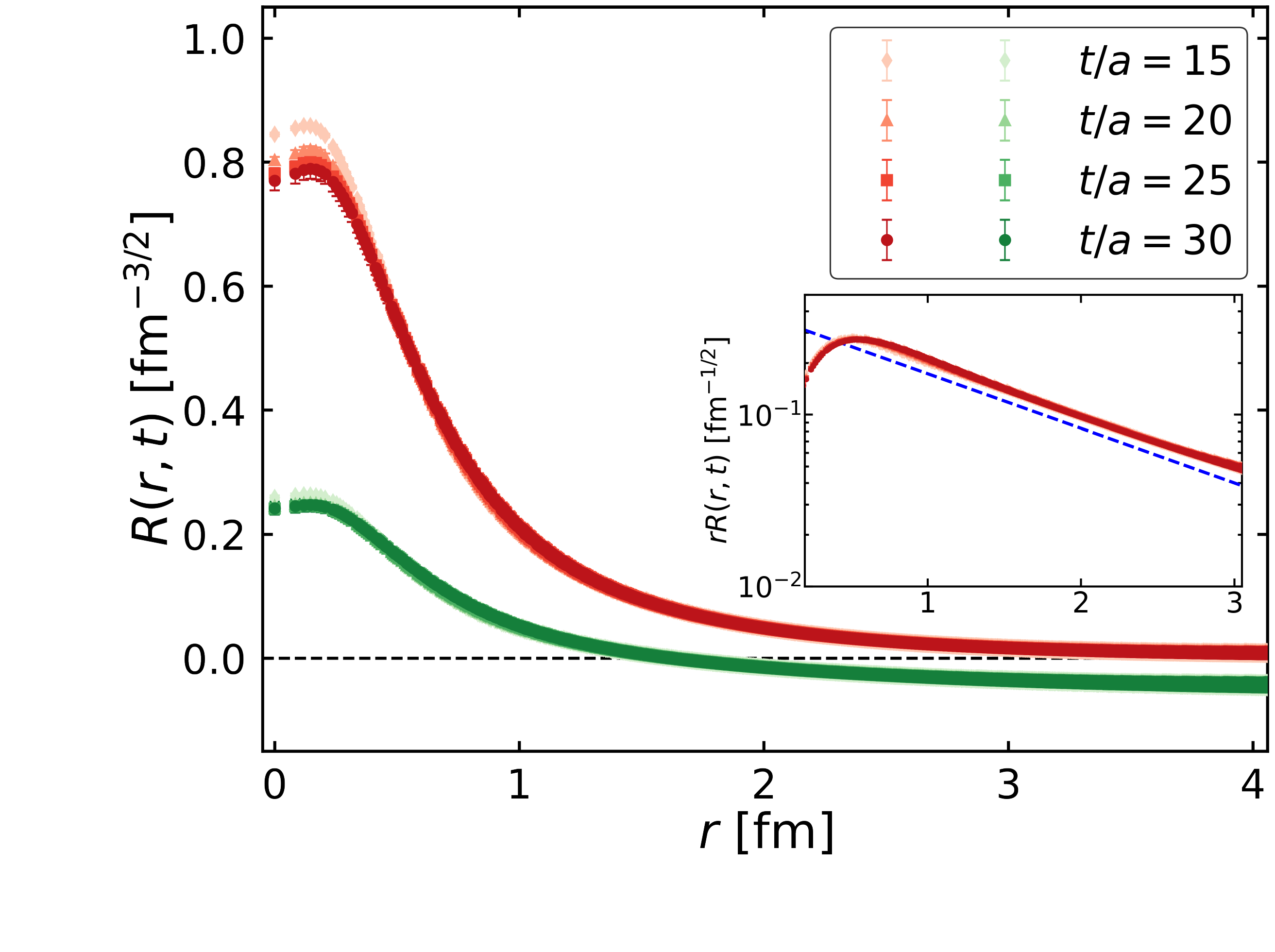}
  \caption{The $R_{0,1}(\vec r, t)$ correlation functions calculated using the optimized source operator for the ground  (red) and  first excited (green) states 
  at multiple Euclidean time $t$.
  The inset shows the asymptotically exponential decreasing of $rR_0(\vec r, t)$, with a dashed line to guide the eye.
  } \label{Fig-R01}
\end{figure}

Putting $\Psi^{(0)}_{0,1}(\vec r)$ into Eq.~\eqref{Eq-Smear-2},
we implement the optimized 
operators at the source for the ground state and the first excited state to calculate $R^{(1)}_{0,1}(\vec r, t)$, shown in Fig.~\ref{Fig-R01}.
The obtained $R^{(1)}_{0,1}(\vec r, t)$ exhibit very stable profiles against $t$ as quantified in Table~\ref{tab-t-dep}.
This stability indicates that the optimized source operators have a strong overlap with the targeted QCD eigenstates and that the iterative procedure has converged at this stage
(the superscript is omitted hereafter for simplicity).
These spatial profiles provide 
a direct characterization of the QCD two-hadron states.
$R_0(\vec r, t)$ displays a localized spatial distribution, characteristic of a bound state.
In contrast, $R_{1}(\vec r, t)$
exhibits an extended profile consistent with a scattering state, with a node appearing  at $r\simeq1.7$ fm.
The asymptotic behavior of $rR_0(\vec r, t)$, shown in the inset, reveals an exponential decrease at large distances, confirming the bound-state nature.
Moreover, the spatial extent of the deviation from the asymptotic behavior indicates that the interaction between two $\Omega_{ccc}$ baryons is confined within $r\lesssim1$ fm.
%==========================
%==========================

\begin{figure}[htbp]
  \centering
  \includegraphics[width=8.cm]{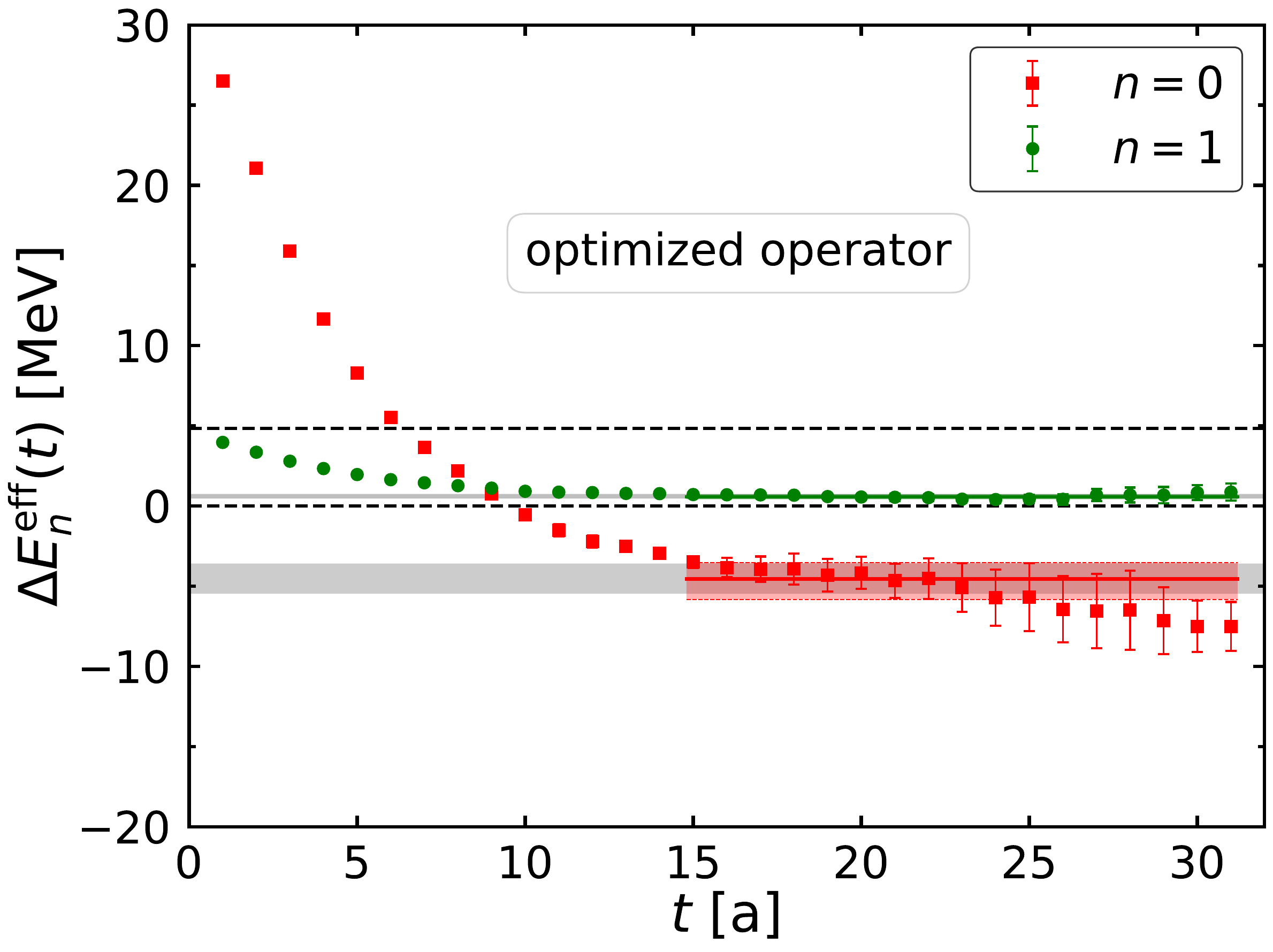}
\includegraphics[width=8.cm]{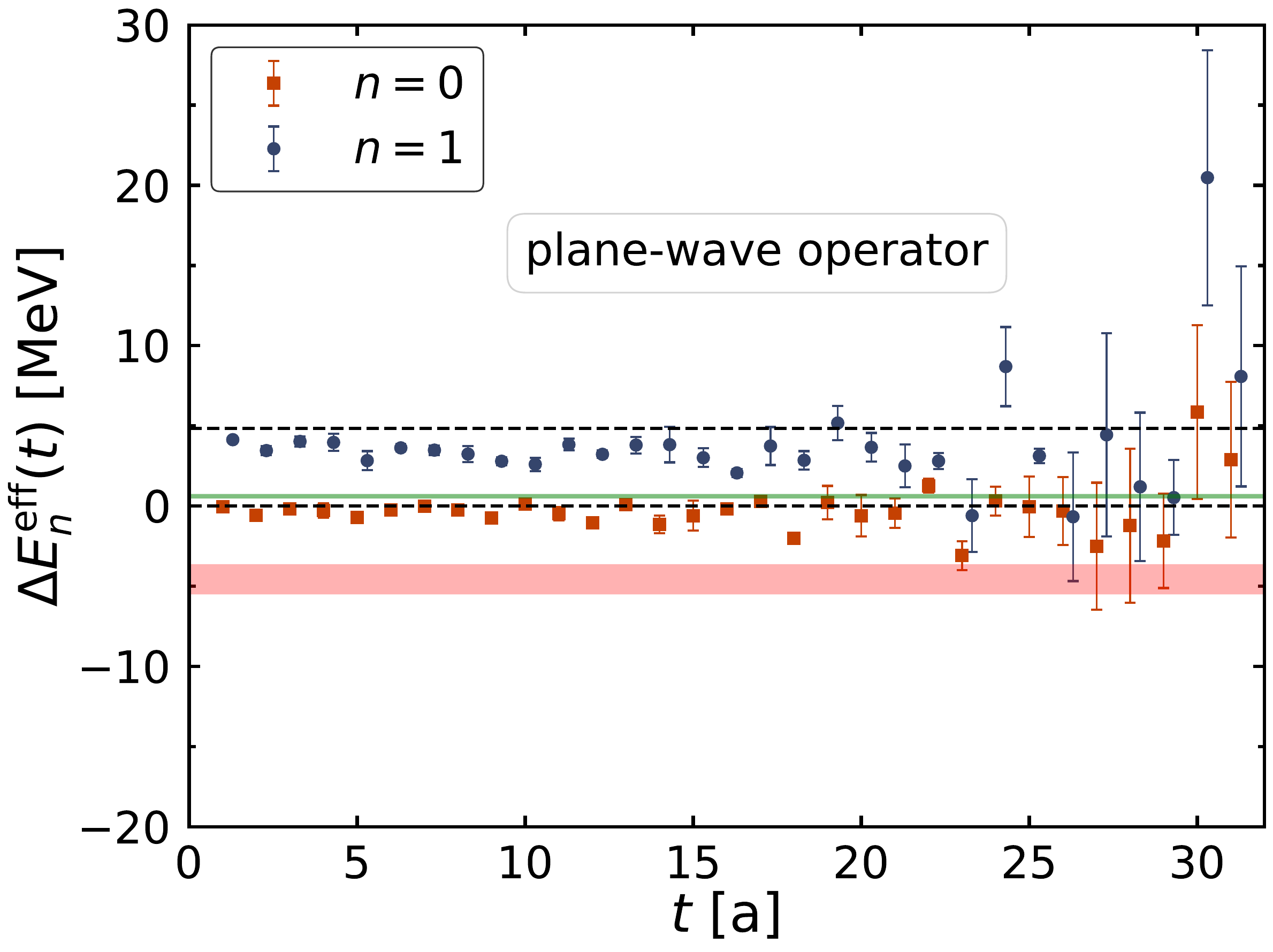}
  \caption{(Top) Effective energies $\Delta E^{\text{eff}}_n(t)=\ln\left[\frac{R_n(t)}{R_n(t+1)}\right]$ from optimized operators for the ground and first excited states, using $R_{0,1}(t)$ defined in Eq.~(\ref{Eq-Rt}). Red and green bands show single-state fits to $R_{0,1}(t)$, consistent with energies from the 3D Schr\"odinger equation shown as gray bands.
  (Bottom) Effective energies from a variational study using two plane-wave operators with relative momenta $p_n=\frac{2\pi}{L}n$, which deviate from the genuine energies. Free spectra are indicated by black dashed lines.
  } \label{Fig-Eeff}
\end{figure}

Shown in Fig.~\ref{Fig-Eeff} (upper) are the effective energies $\Delta E^{\text{eff}}_{0,1}(t)$ defined from $R_{0,1}(t)$ in Eq.~(\ref{Eq-Rt}) obtained by further applying the optimized operators at the sink.
Effective energies $\Delta E^{\text{eff}}_{0,1}(t)$ exhibit stable plateau against a long period of $t$, which are consistent with energies from the Sch\"odinger equation 
calculated using the HAL QCD potentials on a finite box.
These results demonstrate that our method disentangles two states around $2m_{\Omega_{ccc}}\simeq 9700$ MeV, whose energy gap is as narrow as $\sim 5$ MeV. 
Contrary to the naive expectation, 
$\Delta E_0^{\text{eff}}(t)$ shows large statistical errors than $\Delta E_1^{\text{eff}}(t)$. 
Our interpretation is as follows. The ground state wavefunction is more localized than the first excited one. Hence,
the number of space points contributing to the signal in Eq.~\eqref{Eq-sp-src} is effectively less for the ground state than the first excited state.

%==========================
%==========================

%The exceptionally large overlap of the optimized operators to targeted states arises from the highly nontrivial spatial wavefunctions $\Psi_{0,1}(\vec r)$ incorporated in their construction.
To further demonstrate the impact of inter-hadron wavefunctions in operators, 
we performed additional calculations using plane-wave operators, 
corresponding to inter-hadron wavefunctions 
being plane waves 
$\Psi_{0,1}(\vec r)\sim e^{-i\vec p_{0,1}\cdot \vec r}$ projected to $A_1$ irrep with $|\vec p_0|=0$  and $|\vec p_1|=\frac{2\pi}{L}$.
Adopting the same smeared single-baryon operators at both source and sink, we constructed a $2\times2$ Hermitian matrix of temporal correlation functions and extracted the energies via solving a generalized eigenvalue problem (GEVP) following conventional variational analysis~\cite{Luscher:1990ck,Blossier:2009kd}.
The resulting effective energies, shown in Fig.~\ref{Fig-Eeff} (lower), 
exhibit clear deviations from the genuine energies obtained with the optimized operators.
Therefore, we conclude that a variational analysis using only a few plane-wave operators is insufficient to match the performance of the optimized operators.
Based on an estimate from enlarging plane-wave operator set at the sink, a $10\times10$ or larger correlation matrix in GEVP would be required to achieve comparable results.

%==========================
\iffalse
\begin{figure}[htbp]
  \centering
\includegraphics[width=8.cm]{Fig_DEeff_GEVP_p0_p1_smr.pdf}
  \caption{The effective energies obtained from a variational study  using two plane-wave operators with relative momenta $|\vec p_0|=0$ and $|\vec p_1|=\frac{2\pi}{L}$,
  which deviate from the the genuine energies (bands)  determined in Fig.~\ref{Fig-Eeff}.
  } \label{Fig-Eeff-GEVP}
\end{figure}
\fi
%==========================

%=================================================
{\it Summary and outlook.$-$}
In summary, we have presented a systematic framework for constructing optimized two-hadron operators by incorporating realistic hadronic spatial wavefunctions. To this end, we proposed a novel quark smearing technique that utilizes $Z_3$ noise to efficiently implement these operators at source. 
As a test case, we applied this approach to the $\Omega_{ccc}\Omega_{ccc}$ system in the ${}^1S_0$ channel, and successfully constructed highly optimized operators for both the ground and first excited states,
which lead to characteristic spatial structures for the corresponding bound and scattering states.
These operators are shown to outperform combinations comprising only a few plane-wave operators in variational analysis,
and allow for clear state identification even for an energy gap as narrow as $\sim 5$ MeV.

Operator optimization is crucial for accurately determining QCD eigenstates in lattice calculations. 
Our approach is broadly applicable to a wide range of two-particle systems.
%,though those involving quark-annihilation diagrams may require additional consideration.
Applying these operators at the source and sink allows for a better determination of both spectra and matrix elements.
In particular, it opens the door to systematic studies of the nucleon-nucleon systems at the physical point from QCD, thereby providing a fundamental link between nuclear physics and QCD. 
Moreover, the optimized operators constructed in this work offer a superior basis than plane-wave operators for the conventional variational analysis.
%Our findings also provide insight for state-preparations in quantum computing~\cite{Chai:2025qhf}.

%==============================================================================

%=======================================================================================
\begin{acknowledgments}
We thank members of the HAL QCD Collaboration for stimulating discussions. This work was partially supported by RIKEN Incentive Research Project (``Unveiling pion-exchange interactions between hadrons from first-principles lattice QCD"), RIKEN TRIP initiative, the JSPS (Grant Nos. JP22H00129, JP22H04917, JP23H05439, and JP25K17384), and the Japan Science and Technology Agency (JST) as part of Adopting Sustainable Partnerships for Innovative Research Ecosystem (ASPIRE Grant No. JPMJAP2318). The lattice QCD calculations were carried out by using Fugaku supercomputers at RIKEN through HPCI System Research Project (hp240157 and hp240213),  ``Program for Promoting Researches on the Supercomputer Fugaku'' (Simulation for basic science: from fundamental laws of particles to creation of nuclei) and (Simulation for basic science: approaching the new quantum era)
(Grants No. JPMXP1020200105, JPMXP1020230411).
\end{acknowledgments}

\
%\bibliography{Reference}
%=======================================================================================
%apsrev4-2.bst 2019-01-14 (MD) hand-edited version of apsrev4-1.bst
%Control: key (0)
%Control: author (8) initials jnrlst
%Control: editor formatted (1) identically to author
%Control: production of article title (0) allowed
%Control: page (0) single
%Control: year (1) truncated
%Control: production of eprint (0) enabled
%

%=======================================================================================
%=======================================================================================
%\clearpage
%\begin{center}
%\large{\bf{Supplemental Material}}
%\end{center}
%=======================================================================================

%=======================================================================================

\end{document}